\documentclass[a4paper,11pt]{article}
\usepackage{pos, soul, comment, booktabs}
\usepackage[utf8]{inputenc}
\pdfoutput=1

\makeatletter
     {\bgroup\raggedright\small\section*{\refname
        \@mkboth{\MakeUppercase\refname}{\MakeUppercase\refname}}%
      \list{\name{bib\@arabic\c@enumiv}
            \@biblabel{\@arabic\c@enumiv}}%
           {\settowidth\labelwidth{\@biblabel{#1}}%
            \setlength\itemsep{0pt}
            \setlength\parskip{0pt}
            \setlength\parsep{0pt}
            \setlength\partopsep{0pt}
            \leftmargin\labelwidth
            \advance\leftmargin\labelsep
            \@openbib@code
            \usecounter{enumiv}%
            \let\p@enumiv\@empty
            \renewcommand\theenumiv{\@arabic\c@enumiv}}%
      \sloppy\clubpenalty4000\widowpenalty4000%
      \sfcode`\.\@m}
     {\def\@noitemerr
       {\@latex@warning{Empty `thebibliography' environment}}%
      \endlist\egroup}
\makeatother

\title{Gradient flow scale setting with tree-level improvement}

\author*[a]{Christian Schneider}
\author[b]{Anna Hasenfratz}
\author[a]{Oliver Witzel}

\affiliation[a]{Center for Particle Physics Siegen, Theoretische Physik 1, Naturwissenschaftlich-Technische Fakultät,
Universität Siegen, 57068 Siegen, Germany}

\affiliation[b]{Department of Physics, University of Colorado, Boulder, CO 80309, United States}

\emailAdd{christian3.schneider@student.uni-siegen.de}

\abstract{Lattice scales defined using gradient flow are typically very precise, while also easy to calculate. However, different definitions of flows and operators can differ significantly, suggesting possible systematical effects. Using a subset of RBC-UKQCD's 2+1 flavor domain wall fermion and Iwasaki gauge action ensembles, we explore differences between $\sqrt{t_0}$ and $w_0$ gradient flow scales, compare the impact of different operators to define the energy density, and study the effect of using tree-level improvement for the gradient flow. We find that for this set of gauge field ensembles Zeuthen flow with Symanzik operators has the  most consistent approach to the continuum limit and exhibit very small cutoff corrections. Tree-level improvement, traditionally used in step-scaling studies, significantly reduces the spread between different operators, but does not lead to an overall improvement when it comes to reducing cutoff effects for gradient flow scales $\sqrt{t_0}$ or $w_0$.
}

\FullConference{%
 The 39th International Symposium on Lattice Field Theory, LATTICE2022\\
  8th-13th August, 2022
  Bonn, Germany
}

\begin{document}
\maketitle

\section{Introduction}

All lattice quantum chromodynamics (QCD) calculations are subject to cutoff effects. To obtain phenomenologically meaningful results, one needs to take the continuum limit of any physical quantity predicted at finite lattice spacing. This means we need to perform an extrapolation to the renormalization group (RG) fixed point where the lattice spacing $a\to 0$.
Precise determination of lattice scales is therefore essential for such continuum limit extrapolation.  Unfortunately the lattice scale itself also exhibits cutoff effects. Like all physical observables, lattice scales are subject to cutoff effects due to the lattice action. 
In addition, operators used to determine physical observables can also contribute cutoff effects.  The gradient flow lattice scales $\sqrt{t_0}$ \cite{Luscher:2010iy} and $w_0$ \cite{Borsanyi:2012zs}  are increasingly popular in determining the lattice spacing as they are easy to calculate and precise. Both of these approaches rely on the gradient flow transformation \cite{Narayanan:2006rf,Luscher:2009eq}, setting the value of the gradient flow renormalized coupling or its derivative to a given value at the corresponding dimensionless lattice values $t_0/a^2$ and $w_0/a$. Both $\sqrt{t_0}/a$ and $w_0/a$  can exhibit large cutoff effects that originate not only from the lattice action, but also from the flow scheme and the operator used in determining the renormalized coupling. 
In this paper we  study discretization  effects of the gradient flow scales $\sqrt{t_0}$ and $w_0$ and give a practical description of a scale least affected by cutoff effects.
We use a subset of RBC-UKQCD's 2+1 flavor domain-wall fermion (DWF) and Iwasaki gauge field ensembles listed in Table \ref{tab:ensembles} \cite{RBC-UKQCD:2008mhs,RBC:2010qam,RBC:2014ntl,Boyle:2017jwu,Boyle:2018knm}. Specifically we use the set of DWF ensembles generated using the Shamir kernel \cite{Kaplan:1992bt,Shamir:1993zy,Furman:1994ky} and list their key parameters: bare gauge coupling $\beta$, spatial and temporal extent, $L/a$ and $T/a$, input quark mass of the two degenerate light flavors, $am_\ell$, and input quark mass of the strange quark, $am_s$, as well as the number of used configurations. In the following we refer to these ensembles using names indicating their lattice spacing which is related to the inverse of the bare gauge coupling $\beta$. Hence we will use {\it coarse}\/ (C1, C2), {\it medium}\/ (M1, M2, M3), {\it fine}\/ (F1), and {\it extra fine}\/ (X1), where larger integers indicate heavier light quark masses. On these gauge field configurations we perform Wilson (W) and Zeuthen (Z) gradient flow \cite{Ramos:2014kka,Ramos:2015baa} measurements, and determine renormalized couplings using three different operators: clover (C), Wilson (W), and Symanzik (S). In addition we analyze renormalized couplings with and without tree-level normalization (tln) corrections \cite{Fodor:2014cpa}. 

\begin{table}[tb]
  \centering
  \begin{tabular}{cccccccc} 
    \toprule
ensemble   & $\beta$ & $L/a$ & $T/a$ & $am_\ell$ & $am_s^\text{sea}$ & $am_\text{res} $ &$N_\text{cfg}$ \\\midrule
C1 & 2.13 & 24 & 64 & 0.005 & 0.040 & 0.003154(15) & 1636 \\
C2 & 2.13 & 24 & 64 & 0.010 & 0.040 & 0.003154(15) & 1419 \\
\midrule
M1 & 2.25 & 32 & 64 & 0.004 & 0.030 & 0.0006697(34) & 628 \\
M2 & 2.25 & 32 & 64 & 0.006 & 0.030 & 0.0006697(34) & 889 \\
M3 & 2.25 & 32 & 64 & 0.008 & 0.030 & 0.0006697(34) & 544 \\
\midrule
F1 & 2.31 & 48 & 96 & 0.002144 & 0.02144 & 0.0009679(21) & 98 \\
\midrule
X1 & 2.37 & 32 & 64 & 0.0047 & 0.0186 & 0.0006296(58) & 119 \\  
    \bottomrule
  \end{tabular}
  \caption{RBC-UKQCD’s 2+1 flavor Shamir domain-wall fermion and Iwasaki gauge field ensembles \cite{RBC-UKQCD:2008mhs,RBC:2010qam,RBC:2014ntl,Boyle:2017jwu,Boyle:2018knm}.}
  \label{tab:ensembles}
\end{table}

\section{The  \texorpdfstring{$t_0$ and $w_0$}{t0 and w0} scales }
The scale $t_0$ was introduced in Ref.~ \cite{Luscher:2010iy} as the lattice gradient flow time $t_0/a^2$ where $t^2\langle E \rangle=0.3$.  $\langle E \rangle$ is the energy density at flow time $t_0$.   Analogously $w_0$ is defined as the square root of the flow time where $t \frac{d t^2 \langle E \rangle}{d t} = W(t/a^2) = 0.3$ \cite{Borsanyi:2012zs}. These definitions are appropriate to define a lattice scale because the quantity $t^2\langle E \rangle$ has no canonical or anomalous dimension and therefore can be considered as a running renormalized coupling \cite{Hasenfratz:2019puu,Hasenfratz:2019hpg}.  With the normalization $\mathcal{N} =128 \pi^2/3(N_c^2-1)$ the gradient flow coupling $g^2_{GF}= \mathcal{N} t^2\langle E\rangle$ matches the ${\overline{\textrm{MS}}}$ coupling at tree-level.

Restating the definition of the lattice scales, one defines the $t_0$ scale as the lattice gradient flow time $t_0/a^2$ where the renormalized coupling $g^2_{GF}(t_0)=0.3 \mathcal{N}$ and the $w_0$ scale as the lattice flow time $\sqrt{t}/a$ where the renormalization group $\beta$ function is $\beta\left( g^2_{GF}(t) \right)= t \frac{d g^2_{GF}}{d t}= 0.3 \mathcal{N}$. 
\subsection{Tree-level Normalization}

The lattice action induces $\mathcal{O}(a^n)$ cutoff effects at tree-level and, moreover, loop corrections enter $\propto g^{2}\text{log}(a)$. Also the renormalized coupling $g^2_{GF}$ has both tree-level and quantum loop cutoff corrections.  The gradient flow, on the other hand, is a classical operation on the background gauge field configurations\footnote{In Wilsonian renormalization group language the gradient flow defines a renormalization group transformation. In QCD,  non-pathological RG transformations have a fixed point on the $g^2_0=0$ critical surface. An optimally chosen RG transformation has a fixed point and a corresponding renormalized trajectory that are close to the bare lattice action, thus reducing cutoff effects in $g^2_{GF}$ \cite{DeGrand:1995ji,Carosso:2018bmz}}. Following the Symanzik improvement program it is possible to remove all $\mathcal{O}(a^2)$ terms from the flow and operator by choosing Zeuthen flow and Symanzik improved operator to  estimate the energy density $\langle E \rangle$. One can even go further. Tree-level cutoff effects for a given action/flow/operator combination can be expressed as a 4-dimensional integral, or, in finite volume, as a 4-dimensional sum. By redefining the gradient flow  coupling as $g^2_{GF} =\mathcal{N} t^2\langle E(t) \rangle / C(t,L,T)$, where $C(t,L,T)$ is determined analytically according to Eq.~(3.15) of Ref.~\cite{Fodor:2014cpa}, we obtain a tree-level improved (tln) scale. 
Using this method, a notable improvement for the determination of the step-scaling $\beta$ function for SU(3) with $N_f = 4,6,8,10,12$ at sufficiently weak coupling is achieved in \cite{Hasenfratz:2022zsa,Hasenfratz:2022yws,Hasenfratz:2020ess,Hasenfratz:2019dpr,Hasenfratz:2017qyr}. 
 We repeated the analytical calculation of $C(t,L,T)$  for the case of Iwasaki gauge action, Zeuthen or Wilson gradient flow and Wilson plaquette, Symanzik, or clover operator on $24^3\times 64$, $32^3\times 64$, and $48^3\times 96$ volumes. As a shorthand to refer to the different combinations, we use the convention [action][flow][operator], with the prefix $n$ to denote tln improvement.

Tree-level normalization does not mean all cutoff effects introduced by the lattice action and the GF operator are removed.  Step scaling function calculations that compare different flows, operators, with and without tree-level normalization show that perturbatively improved combinations do not always lead to smaller cutoff effects at strong coupling \cite{Hasenfratz:2019dpr}.

\section{Analysis}

\subsection{Determination of \texorpdfstring{$\sqrt{t_0}$ and $w_0$}{sqrt(t0) and w0}}

\begin{centering}
\begin{figure}[t]
  \centering
  \includegraphics[width=0.48\textwidth]{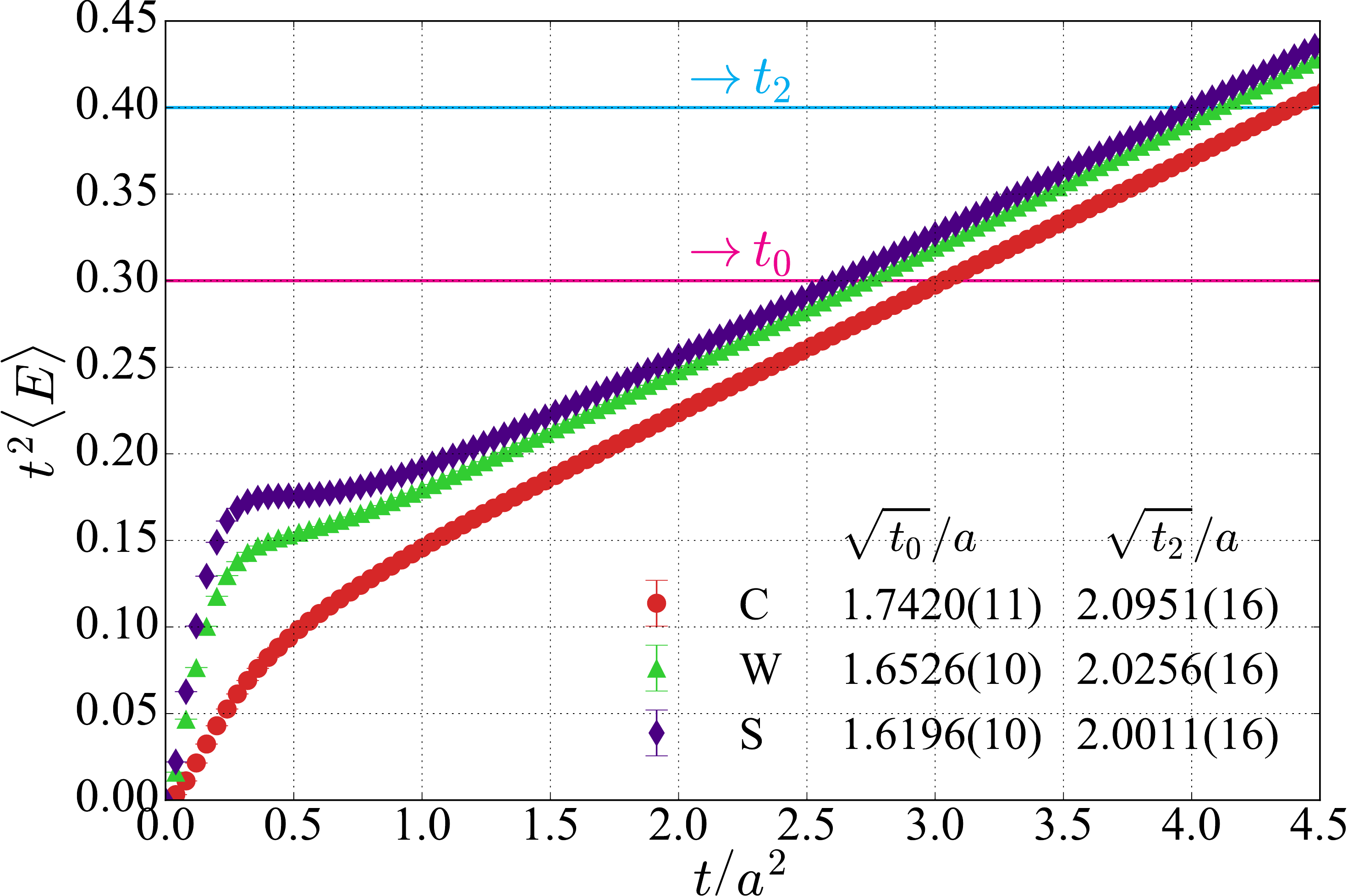}
  \hfill
  \includegraphics[width=0.48\textwidth]{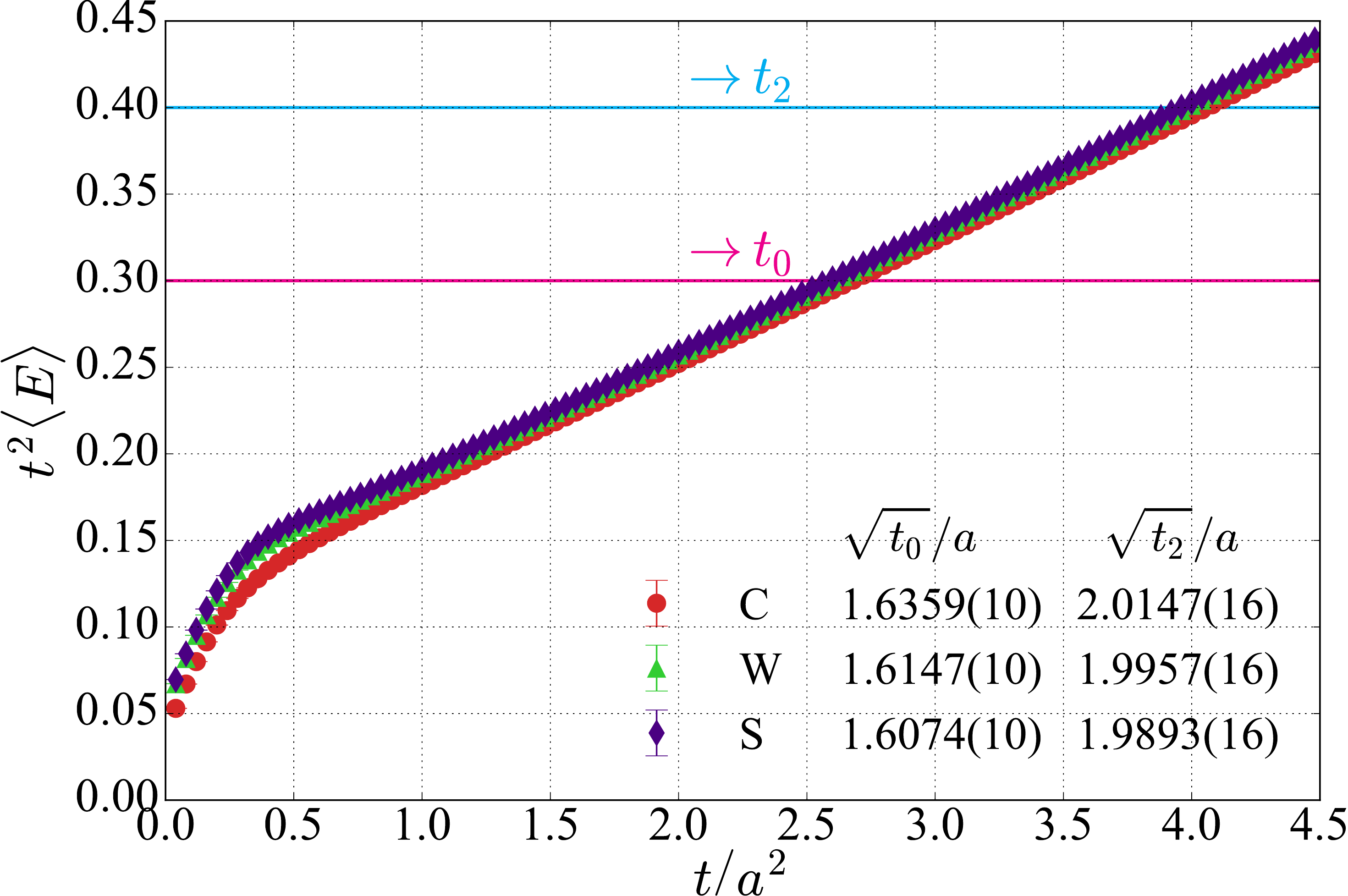}
  \includegraphics[width=0.48\textwidth]{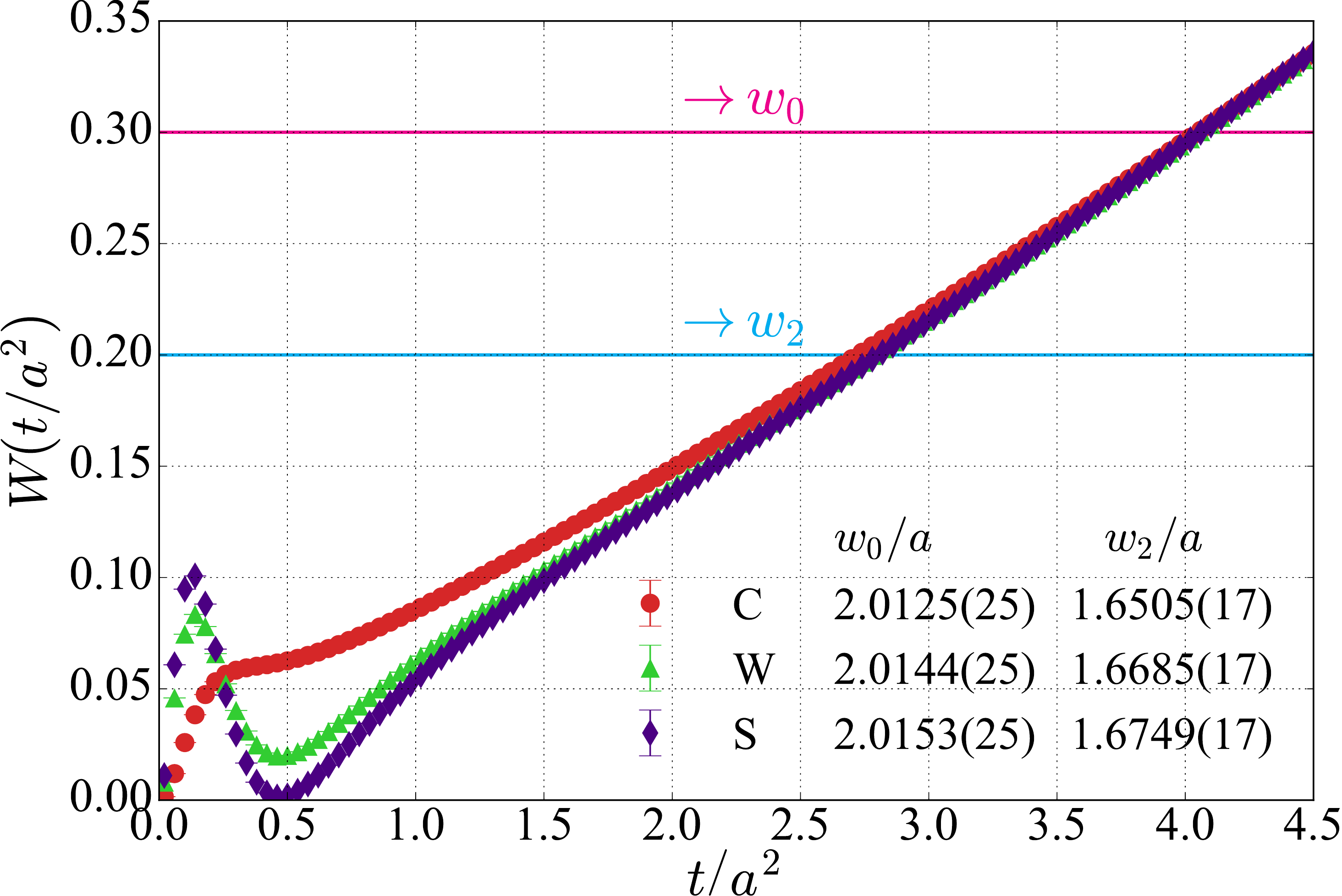}
  \hfill
  \includegraphics[width=0.48\textwidth]{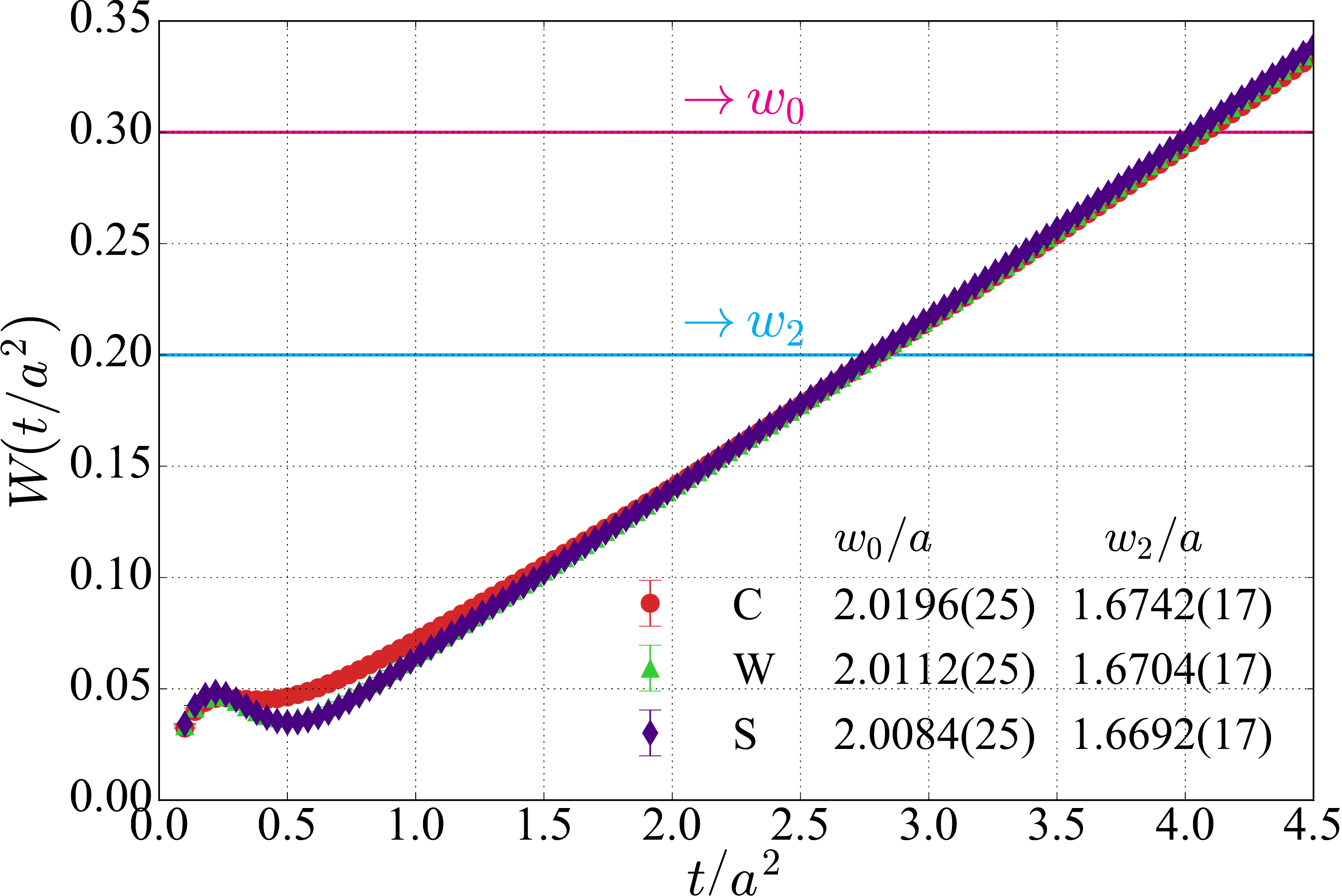}
\caption{Determination of the Wilson flow $t$ scales (top) and $w$ scales (bottom) on the M1 ensemble. Plots on the left show the determination without tln, plots on the right with tln. In addition to the standard $t_0$ and $w_0$ scales we also consider the $t_2$ and $w_2$ scales that are defined at different values of $t^2\langle E \rangle$ and  $W(t/a^2)$.}
\label{fig:gfscales}
\end{figure}
\end{centering}

The scale $t_0$ is defined as the lattice scale $t_0/a^2$ where the dimensionless quantity $g^2_{GF}(t)=\mathcal{N} t^2\langle E \rangle=0.3\mathcal{N}$  \cite{Luscher:2010iy}. The choice of the constant is arbitrary, and in the following we will consider both $t_0$ and a new scale $t_2$ defined where  $g^2_{GF}(t)=0.4\mathcal{N}$. Analogously $w_0$ is defined as the square root of the flow time where $\mathcal{N} t \frac{d t^2 \langle E \rangle}{d t} = W(t/a^2)\mathcal{N} = 0.3\mathcal{N}$ \cite{Borsanyi:2012zs}. In addition we consider the scale $w_2$ defined by $W(t/a^2) = 0.2$.

Figure \ref{fig:gfscales} illustrates the determination of the $t$ and $w$ scales with Wilson flow on the M1 ensemble. The $t$ scales without tln correction show strong dependence on the choice of the operator, whereas tln improvement removes most operator dependence. The $w$ scales show overall better consistency.

\begin{figure}[t]
  \centering
    \includegraphics[width=\textwidth]{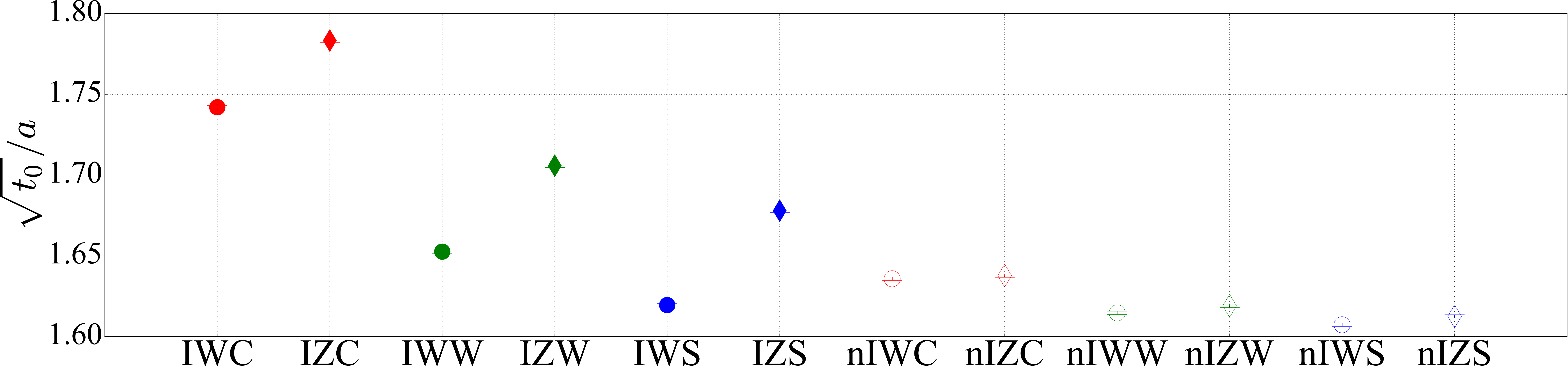}
    \caption{Comparison of $\sqrt{t_0}/a$ values determined for different operators and flows on the M1 ensemble. We use the short-hand [action][flow][operator] indicated by the corresponding first letter to refer to the different combination and prefix a `n' when using tln. }
    \label{fig:t0_comparison}
  \end{figure}

\subsection{Identifying cut-off effects}

\begin{figure}[t]
  \centering
  \includegraphics[width=0.48\textwidth]{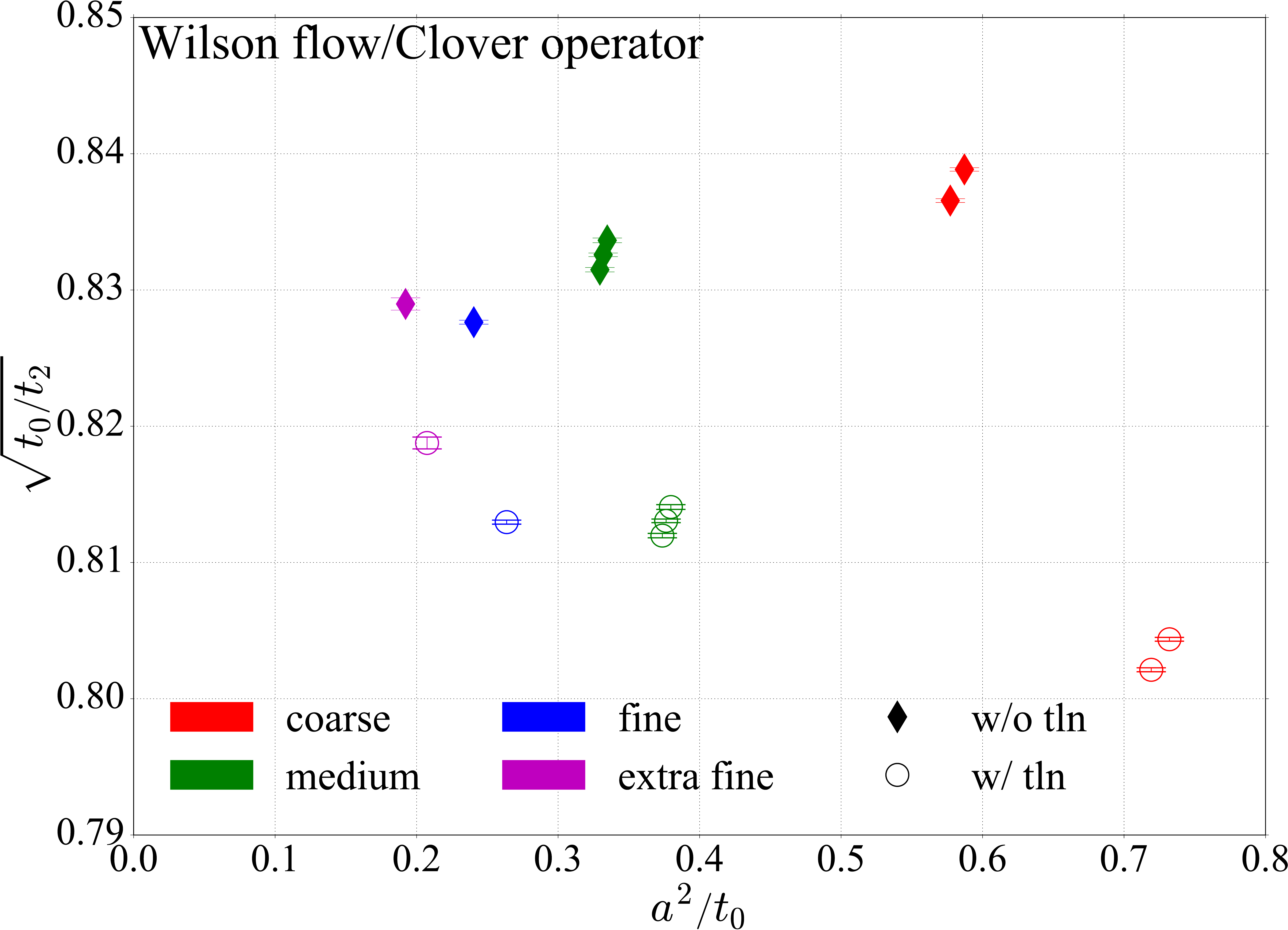}
  \hfill
  \includegraphics[width=0.48\textwidth]{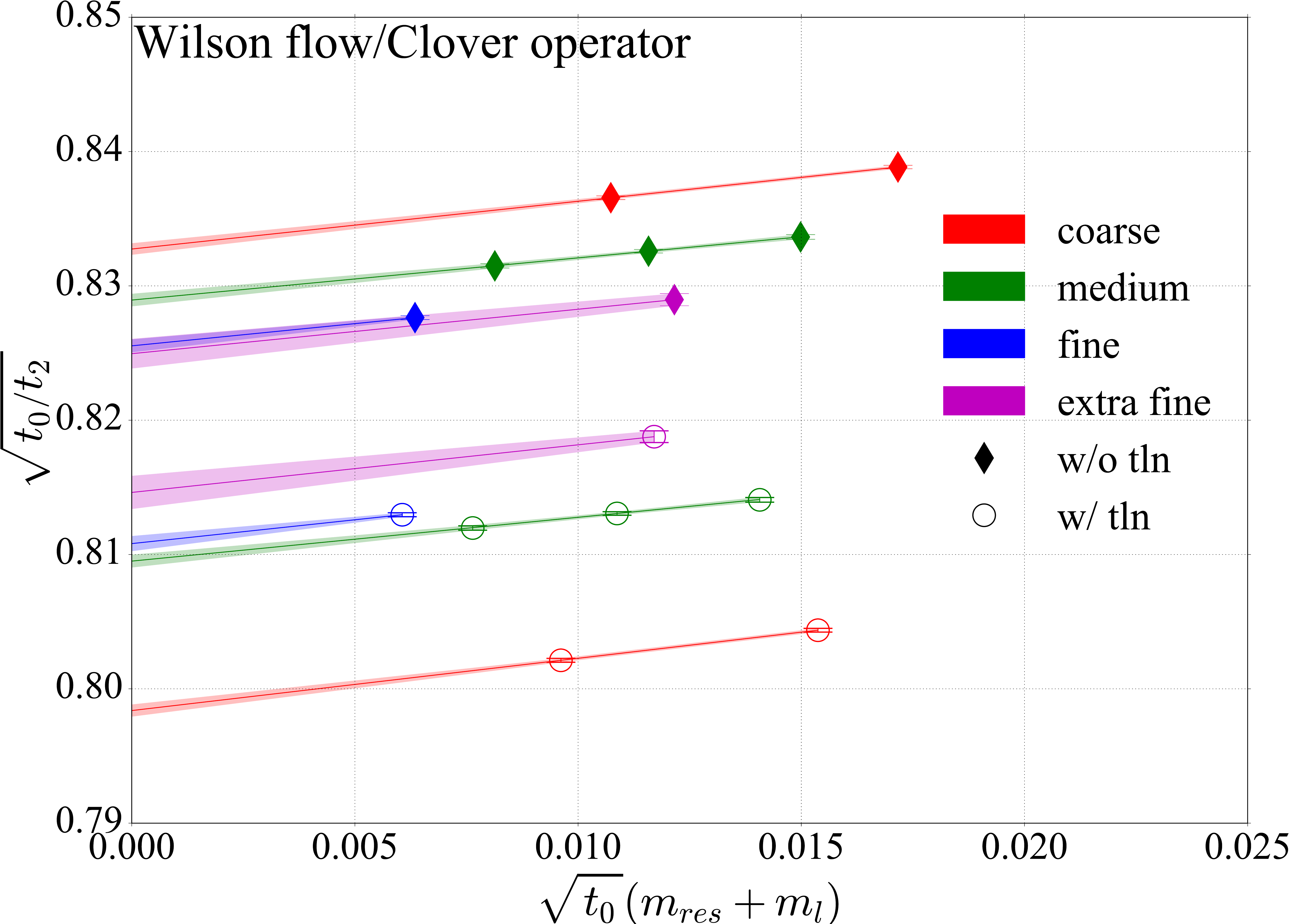}
  \caption{The plot on the left shows the ratio $\sqrt{t_0\left/t_2\right.}$ vs.~$a^2\left/t_0\right.$ using Wilson flow and the clover operator. The dependence on the light sea-quark mass is clearly resolved. We therefore perform a chiral extrapolation in $\sqrt{t_0}(m_\text{res}+m_\ell)$ as shown on the left. The extrapolations of the fine and extra fine ensembles is guided using the slopes of the extrapolations of the coarse and the medium ensembles.}
  \label{fig:chiral_extrapol}
\end{figure}

In Fig.~\ref{fig:t0_comparison} we compare $\sqrt{t_0}/a$ predicted by both Wilson and Zeuthen flow using all three operators with and without tln. 
This plot on its own is insufficient to deduce which flow/operator combination has the smallest cutoff effects
  because a priori the lattice spacing of an ensemble is not known. However, the more than 10\% difference between the predicted scales should serve as a warning. This level of uncertainty could make tuning of, e.g.~quark masses, troublesome.
 To get a handle on the cutoff effects, we  form ratios of scales, e.g.~$\sqrt{t_0/t_2}$.\footnote{A similar idea was presented by Alberto Ramos at this conference \cite{Ramos:2022lat}.} With a given flow and operator combination,  $t_0$ and $t_2$ have similar cutoff corrections, though $t_2$ should have smaller corrections as it is defined at a larger flow time. Thus the ratio $\sqrt{t_0/t_2}$  shows the cutoff effects of the $t_0$ scale relative to $t_2$. We investigate it  as  a function of $a^2/t_0$, assuming that  cutoff corrections on  $a^2/t_0$ have only a small effect.

The left panel of Figure \ref{fig:chiral_extrapol} shows this ratio obtained with  Wilson flow and clover operator, with and without tln, for all investigated ensembles. Symbols with the same color refer to ensembles with the same bare gauge coupling but different masses. This plot demonstrates that the mass dependence is not negligible. If we want to identify cutoff effects, we first need to extrapolate  our data to the chiral limit. The expected mass dependence of the $t_0$ scale is linear in $am$. That is indeed satisfied by our data, as can be observed in the right panel of Fig.~\ref{fig:chiral_extrapol} where we extrapolate the ratio $\sqrt{t_0/t_2}$ linearly in $\sqrt{t_0} m_f=\sqrt{t_0}(m_\text{res}+m_\ell)$ for the C and M ensembles. 
For the F and X ensemble, we only have data at one value of the light quark mass $m_\ell$. Observing that the slopes of the C and M ensembles are similar, we take the slopes of the C and M ensembles to predict the ratio $\sqrt{t_0/t_2}$ in the chiral limit and use the differences in the slopes to inflate the uncertainty. At this point we have chiral limit values for our scales for each of the four different bare gauge couplings.

\begin{figure}[t]
  \centering
  \includegraphics[width=0.48\textwidth]{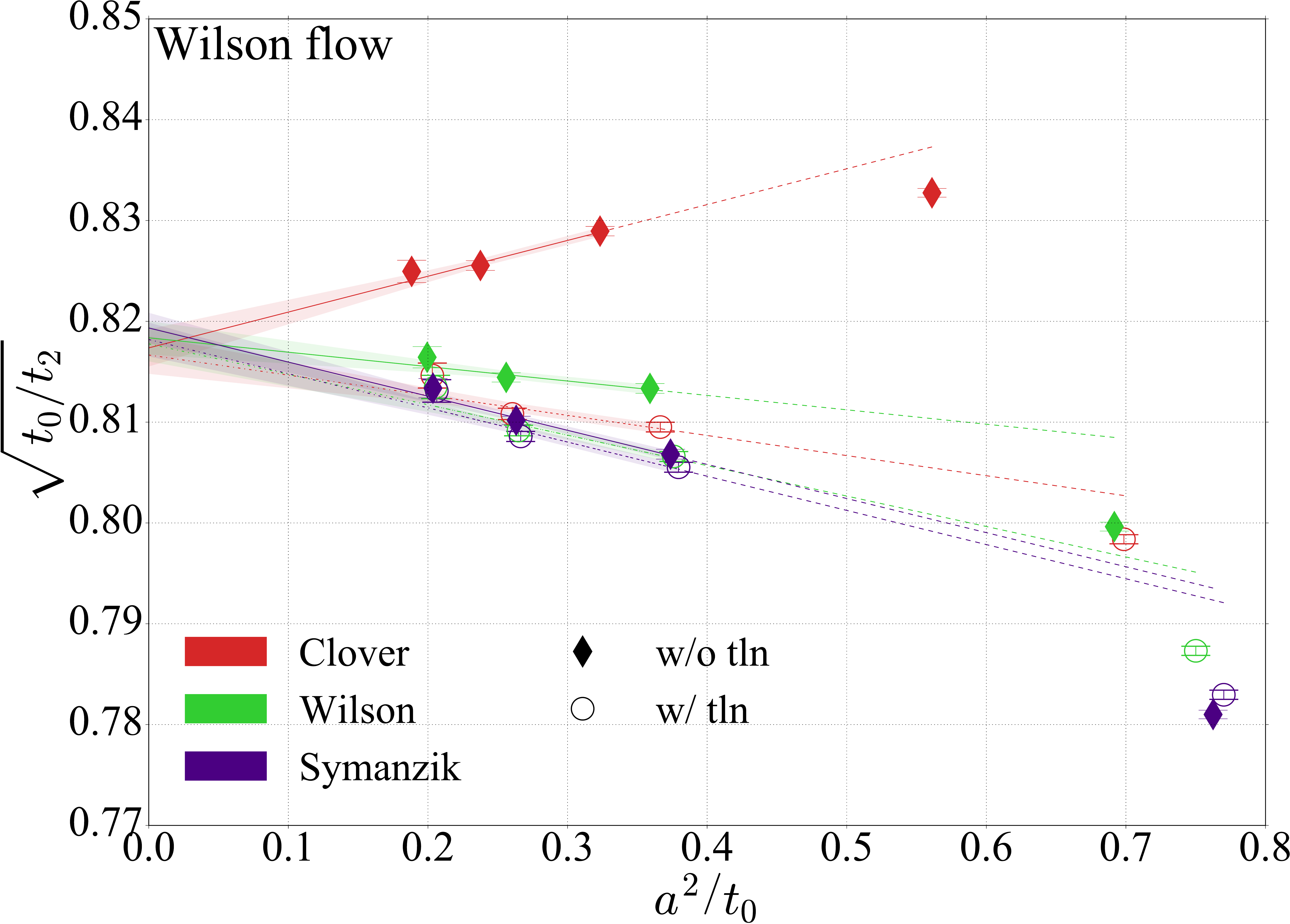}
  \hfill
  \includegraphics[width=0.48\textwidth]{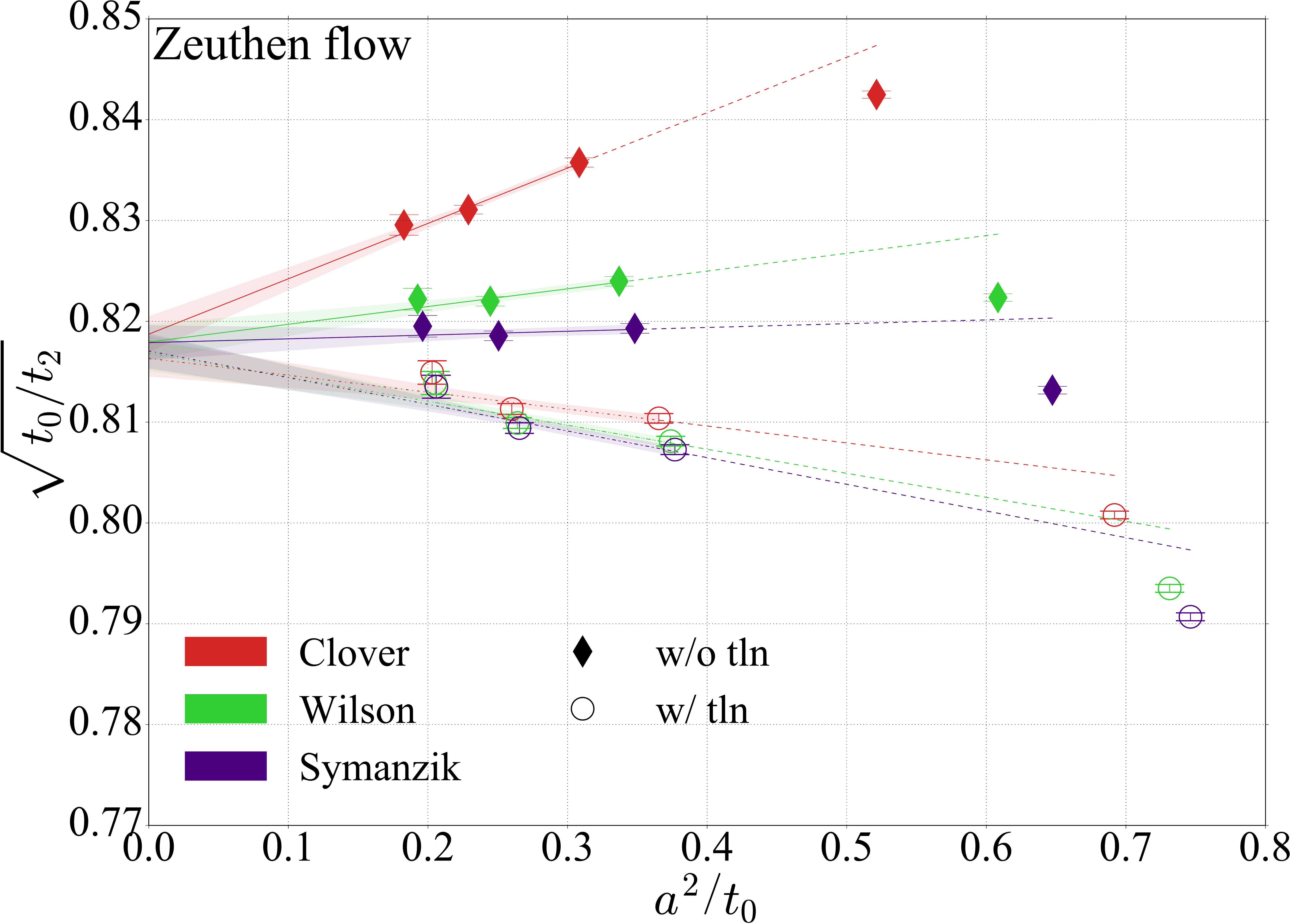}\\[2mm]
  \includegraphics[width=0.48\textwidth]{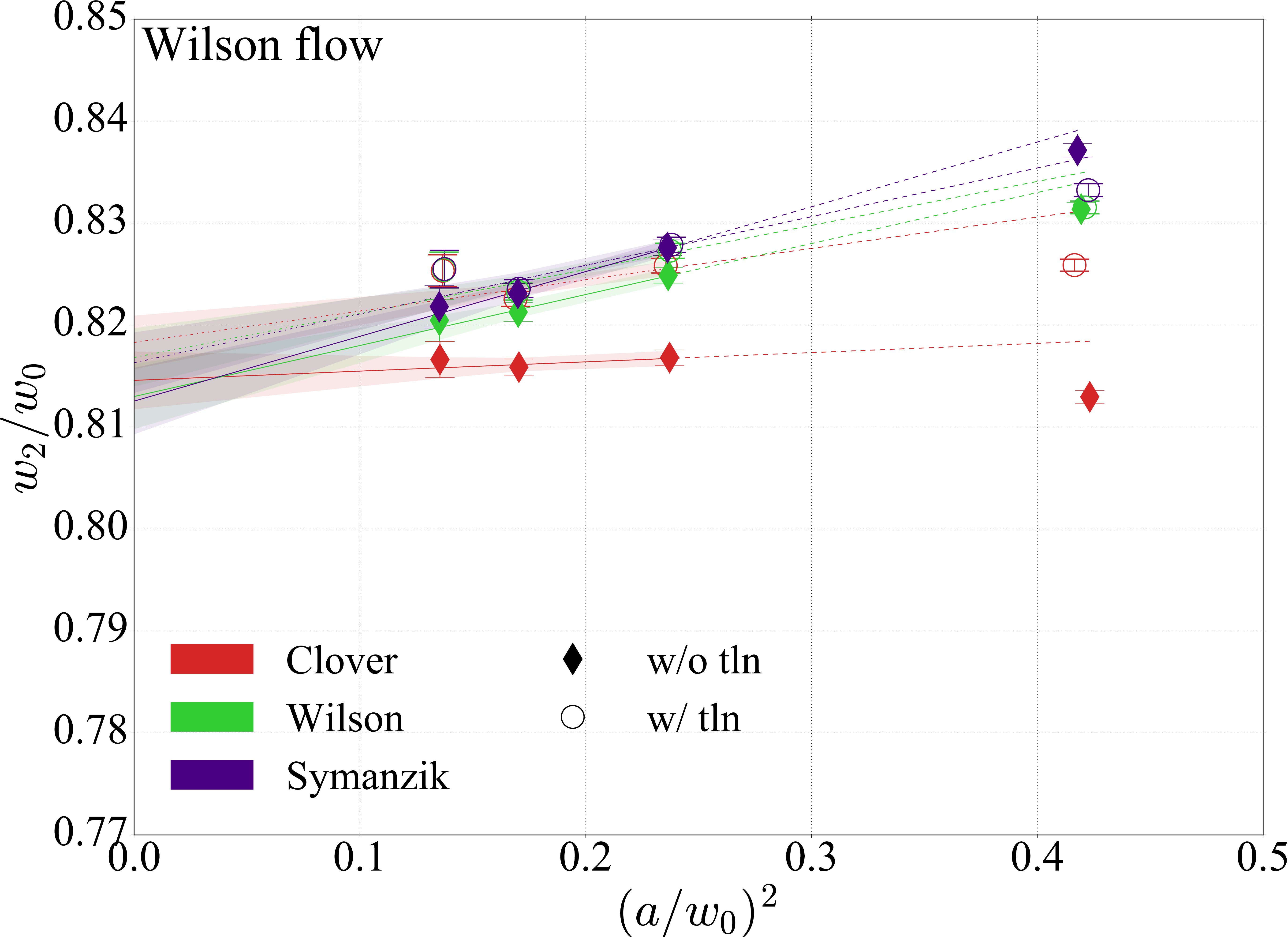}
  \hfill
  \includegraphics[width=0.48\textwidth]{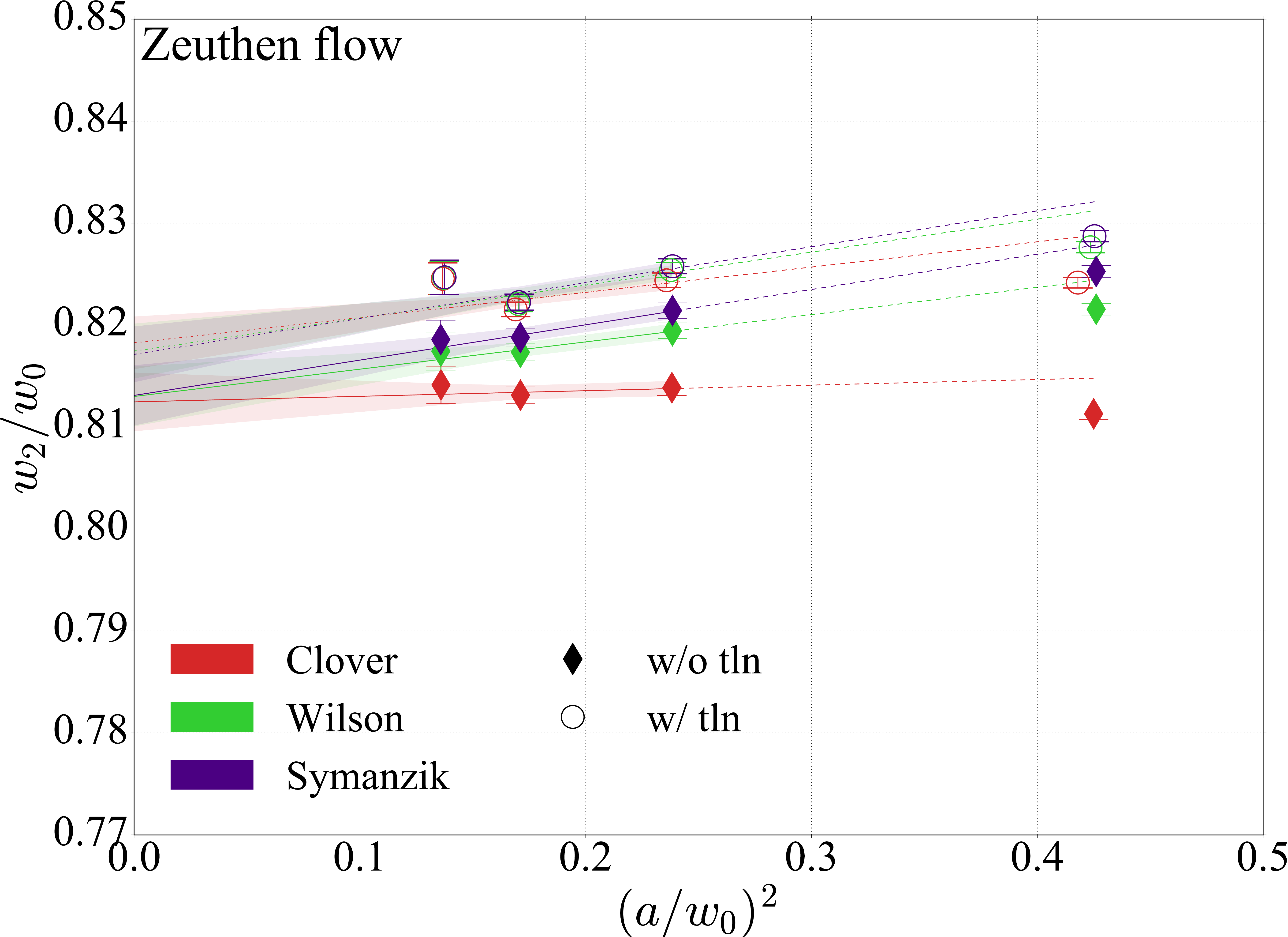}\\[2mm]
  \includegraphics[width=0.48\textwidth]{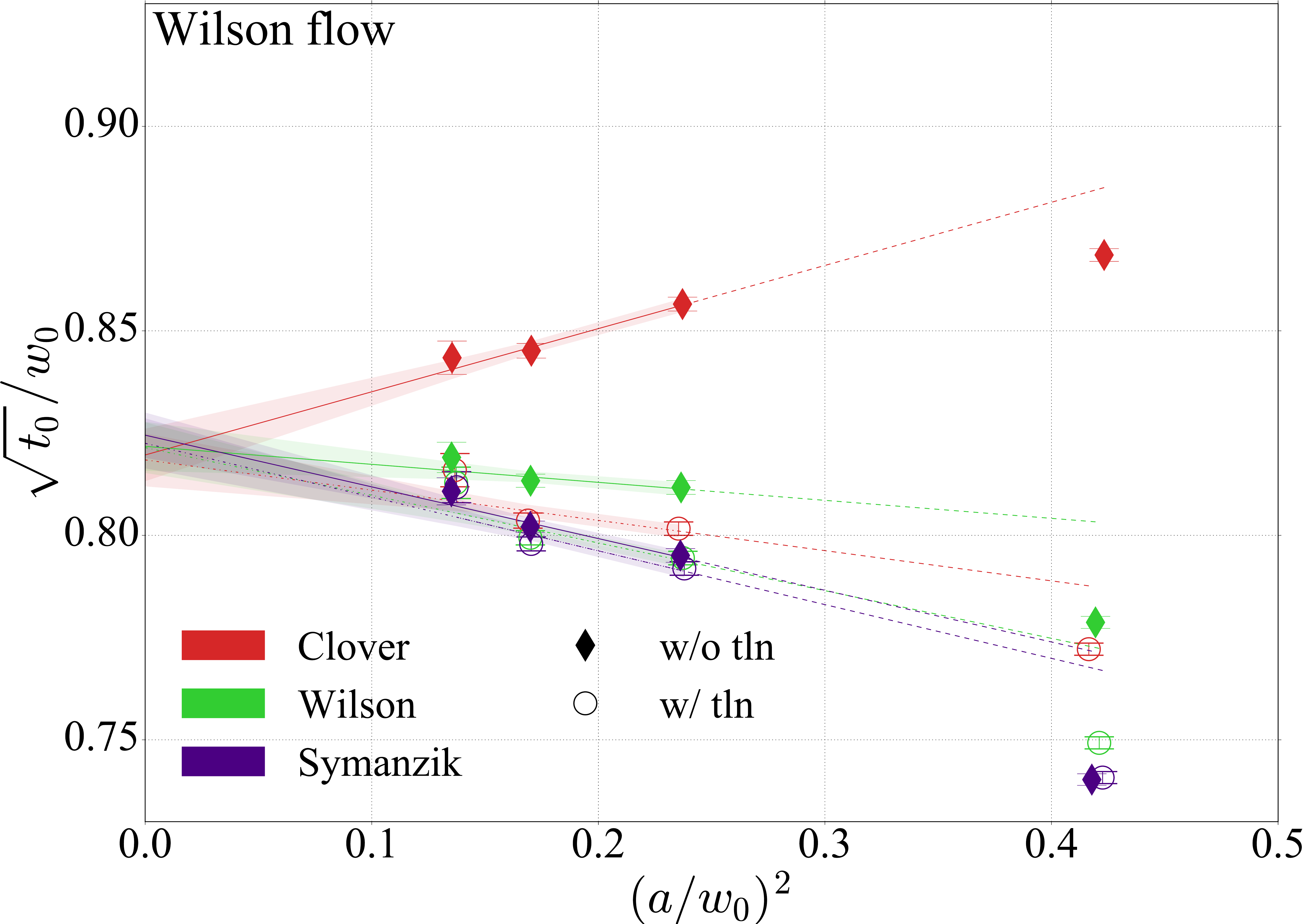}
  \hfill
  \includegraphics[width=0.48\textwidth]{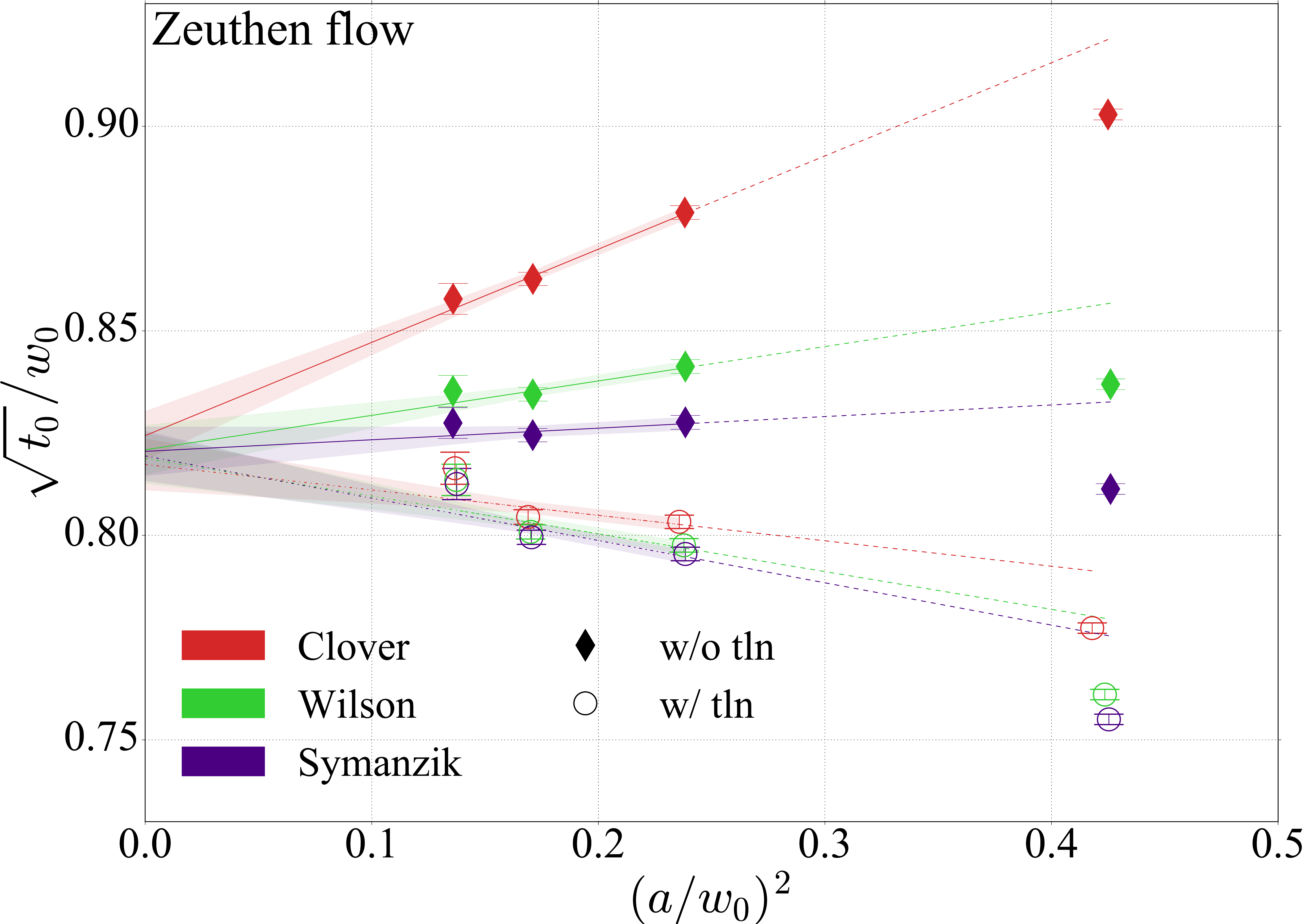}
\caption{Continuum extrapolation of the chirally extrapolated data for different ratios using chiral extrapolations with $t$ or $w$-scaled $x$-axis. The coarse ensembles are not included in the extrapolation fits since our simple, ``linear-in-$a^2$'' fit ansatz is insufficient to describe them. Higher order corrections are likely needed for these ensembles.}
\label{fig:continuum_extrapols}
\end{figure}

Finally we can explore the continuum limit by examining the ratios in the chiral limit. In the top panels of Figure \ref{fig:continuum_extrapols} we study continuum limit extrapolations for $\sqrt{t_0/t_2}$ showing data based on Wilson flow on the left and Zeuthen flow on the right. In all cases we observe that the data points corresponding to the coarse ensembles are not described by the leading order $\mathcal{O}(a^2)$ form. The right most data points are therefore not included in the continuum fit indicated by the solid lines with shaded error band. To visualize the tension with the coarsest data points, we continue the fit lines to the right using dotted lines. Different flow/operator combinations have different approaches to the continuum but should have the same continuum limit value. Also the left most data point corresponding to the extra fine ensemble seems to be slightly off. That value may suffer from the fact that the X1 ensemble has poor statistical properties (e.g.~almost frozen tunneling of the topological charge) and, moreover, due to its large light input quark mass, requires a long chiral extrapolation. Looking at the $p$-values to study the goodness of our continuum extrapolation fits, we observe $p$-values $\gtrsim 20\%$ for fitting data without tln but $p$-values drop to around $3-4\%$ for tln improved data. Also our other ratios exhibit significantly lower $p$-values for continuum extrapolations of tln improved data compared to fits of unimproved data. In all cases we find that the data point derived from the extra fine ensemble has an increased contribution to the total $\chi^2$ raising further concerns about X1.

The ratio $\sqrt{t_0/t_2}$ shows for both flows less than 2\% cutoff effects on the X, F and M ensembles, but up to 5\% on the coarse ensemble. Interestingly the combination of Zeuthen flow with Symanzik operator exhibits practically no lattice artifacts on the three finer ensembles. We can interpret this as the best combination to predict the lattice scale with Iwasaki gauge action.  Our continuum limit prediction using the IZS combination is $\sqrt{t_0/t_2}=0.8187(18)$, which is consistent with determinations based on other combinations.

The $w$ scales show a similar behavior, as is illustrated in the middle panels of 
Fig.~\ref{fig:continuum_extrapols}, and we again observe that fits to tln improved data have lower $p$-values ($4-5\%$) compared to fitting data without tln ($p$-values $\gtrsim 50\%$). 
The deviation between the different combinations is smaller than for the $t_0$ scale, and the ``optimal'' combination now  appears to be Zeuthen flow with clover
  operator.\footnote{The cutoff effects using $w$ scale appear to be the opposite  of the $t$ scale. This only reflects the definition of $t_2>t_0$,  whereas $w_0>w_2$.}
The continuum limit of the ratio $\sqrt{t_0}/w_0$ has been predicted by several lattice calculations. We  demonstrate its cutoff corrections  in the bottom panels of Fig.~\ref{fig:continuum_extrapols}, where we again observe noticeably lower $p$ values when fitting tln improved data compared to fitting the data without tln.

\begin{figure}[t]
  \centering
  \includegraphics[width=0.44\textwidth]{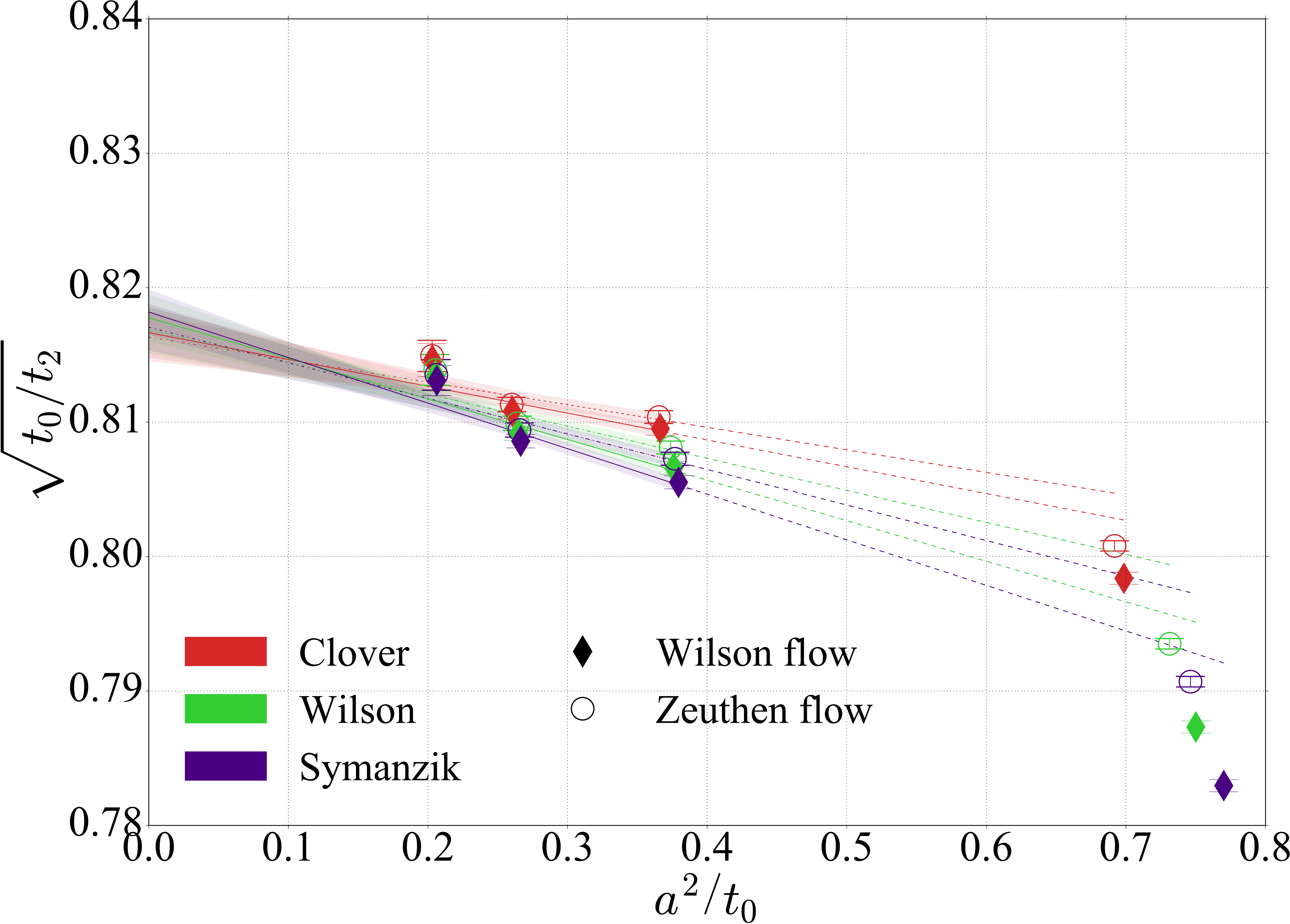}
  \hfill
  \includegraphics[width=0.44\textwidth]{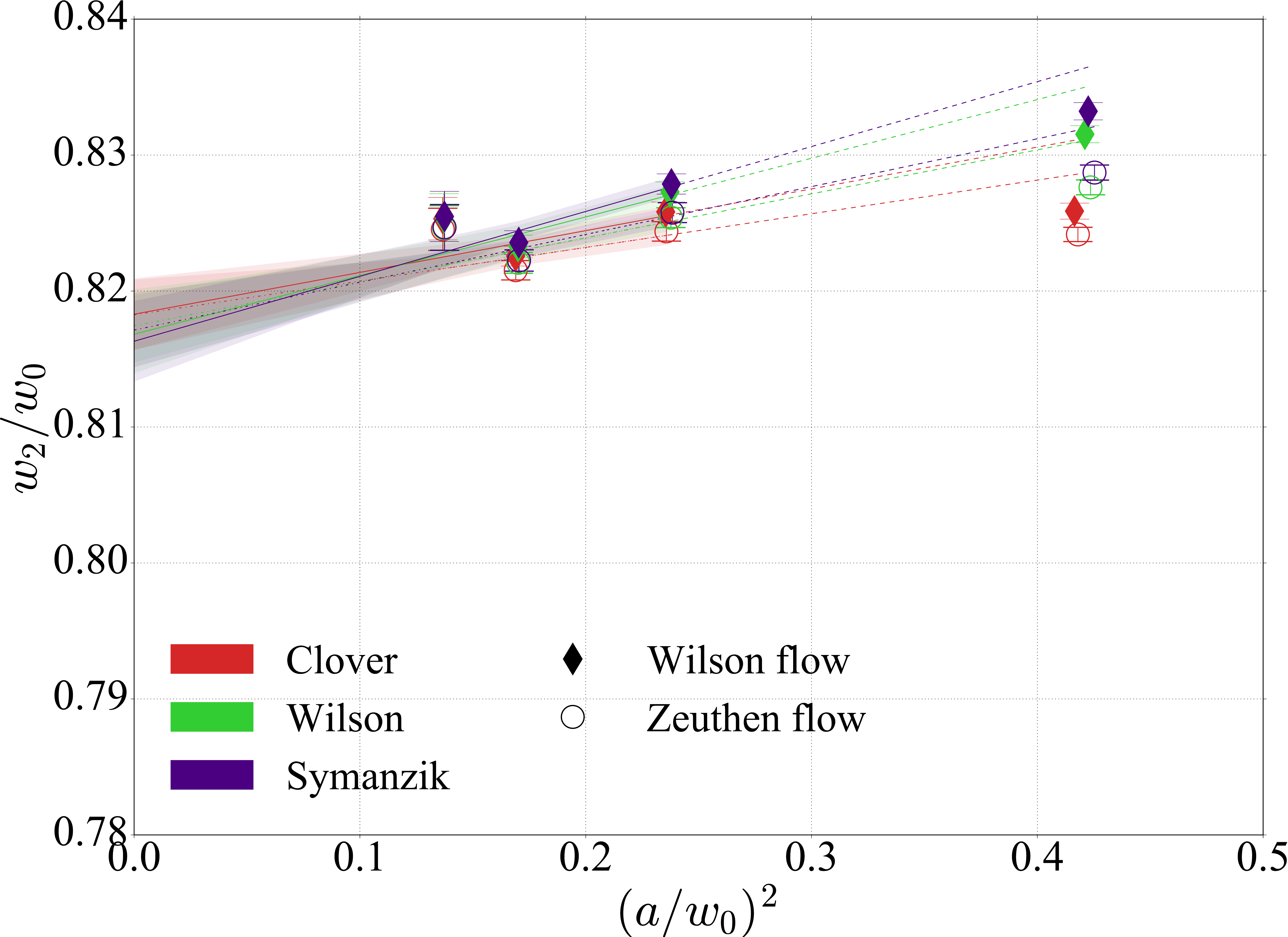}\\
  \includegraphics[width=0.44\textwidth]{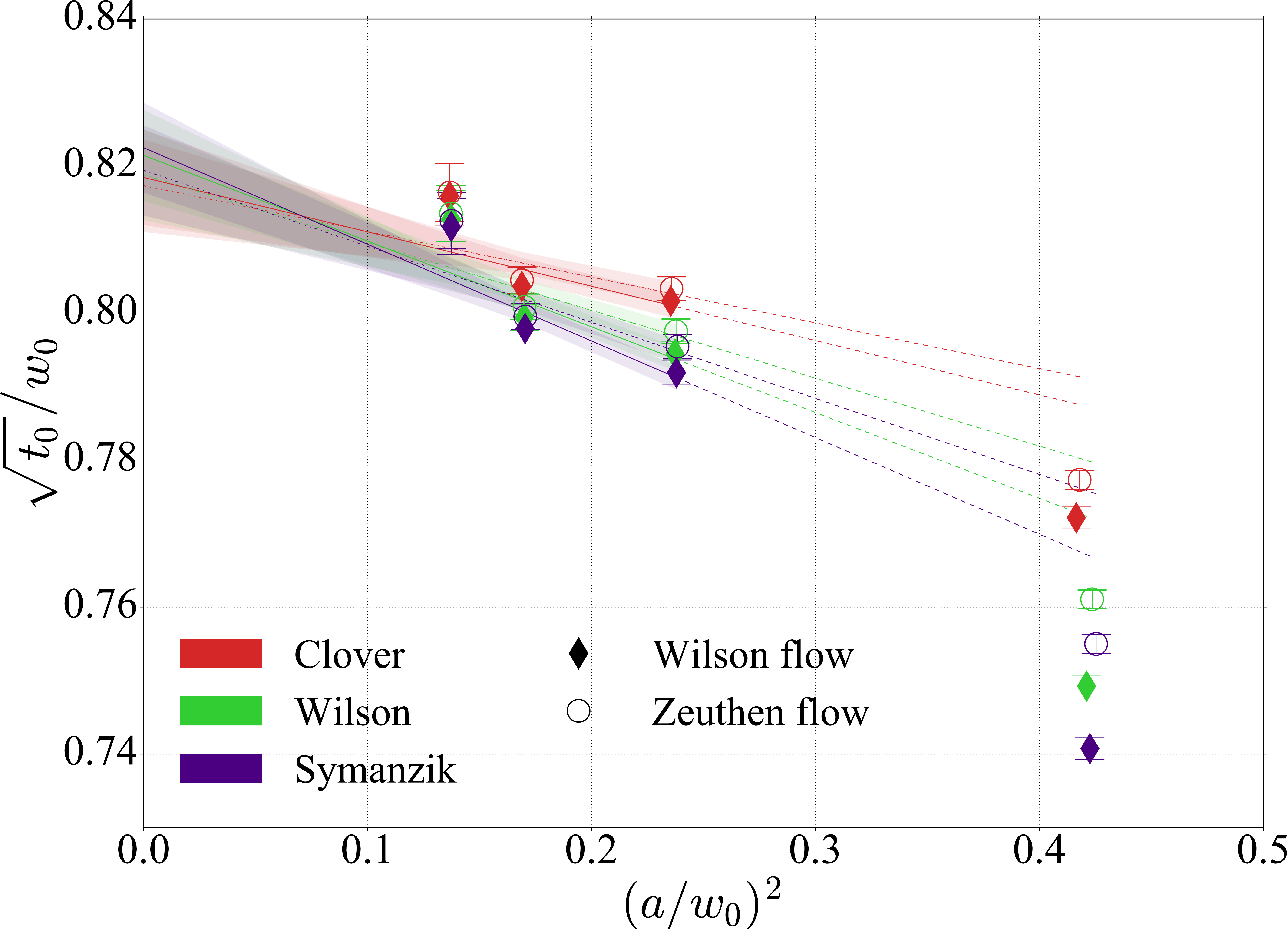}
  \caption{Continuum extrapolations of tln improved ratios of $t$ and $w$ scales. Applying tln largely removes differences observed for different operators and also the differences between Wilson and Zeuthen flow are reduced. However, the overall cutoff effects are not reduced by tln corrections.}
  \label{fig:Zflow_Wflow_tln_comparison}
\end{figure}

We can gain further understanding of the cutoff effects by  focusing at the tree-level improved data gathered in Fig.~\ref{fig:Zflow_Wflow_tln_comparison}. For all three ratios,
we observe close agreement between the tln improved data, independent of the flow and operator considered. However, the tln combinations show significant cutoff effects of approximately $1-2\%$. This is contrary to our expectations, because we expected that removing tree-level cutoff effects will improve the overall scaling behavior. Apparently quantum loop effects  at strong coupling interfere and introduce significant lattice artifacts.

\section{Conclusions and Outlook}

\begin{figure}[t]
  \centering
  \begin{picture}(150,20)
      \put(0,0){\includegraphics[height=0.08125\textheight]{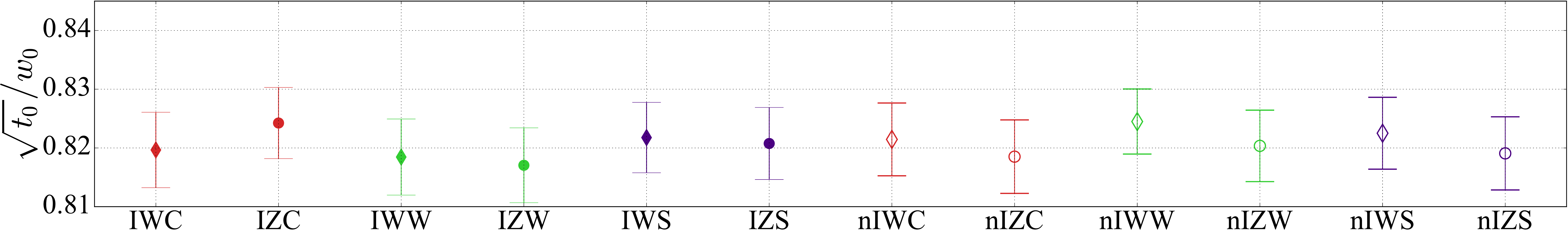}}
      \put(125,0){\includegraphics[height=0.08125\textheight]{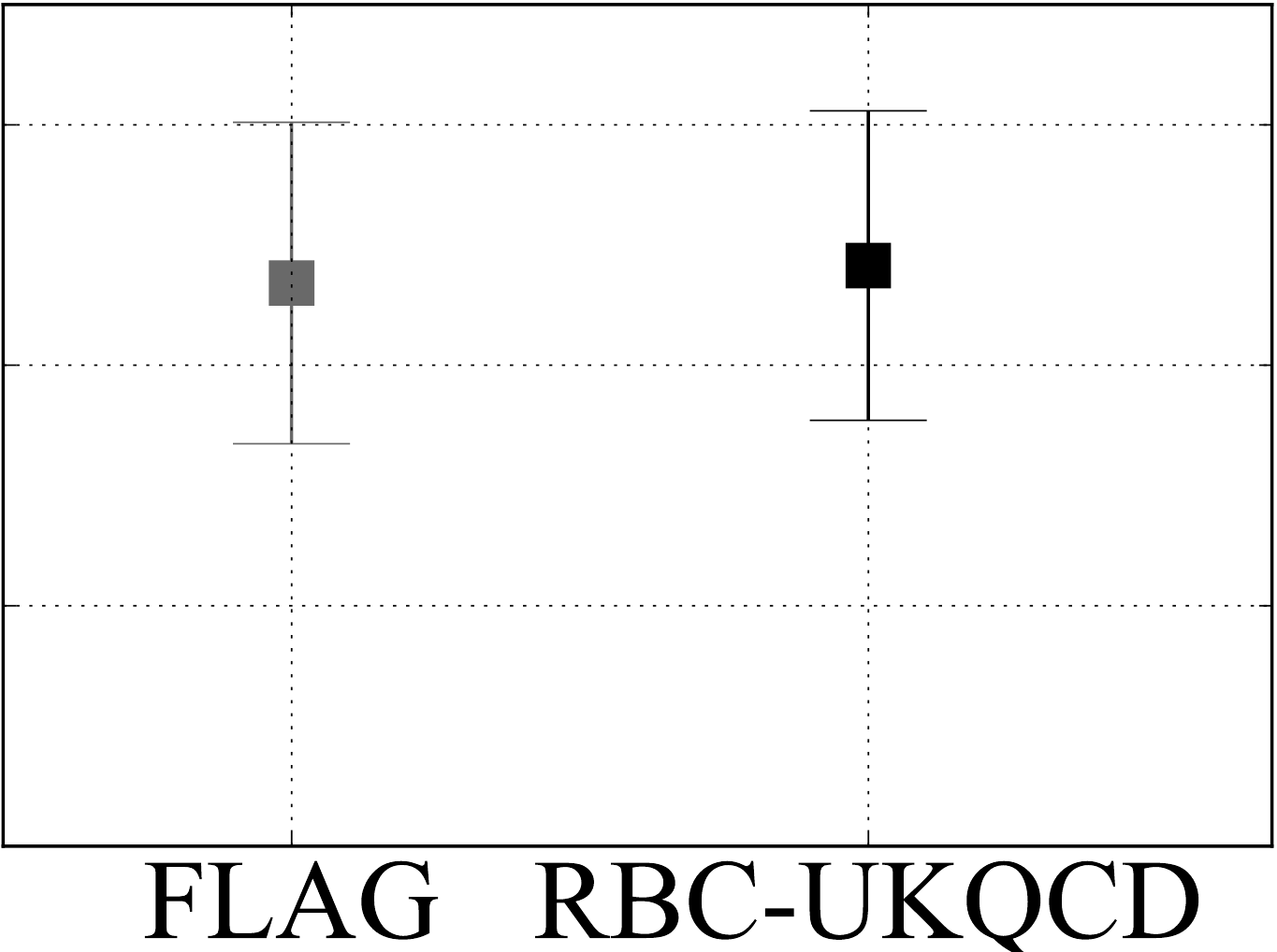}}
      \put(60,15){\scriptsize{chiral limit}}      
      \put(129,5){\scriptsize{physical point}}
  \end{picture}
\caption{The left panel shows a comparison of continuum limit values for $\sqrt{t_0}/w_0$ obtained for different flows/operators in the chiral limit. In addition we show in the small panel on the right ratios at the physical point derived using FLAG 2021 averages \cite{FlavourLatticeAveragingGroupFLAG:2021npn} for 2+1 flavor simulations as well as results by RBC-UKQCD \cite{RBC:2014ntl}, which are based on a subset of the Shamir DWF ensembles used here but include in addition e.g.~the M\"obius DWF ensembles at the physical point (C0, M0).}
\label{fig:FLAG_comparison}
\end{figure}

In this paper we consider the important issue of lattice scale setting. Although gradient flow scales can be determined with high statistical precision,  they may carry significant cutoff effects. It is essential to identify the most reliable gradient flow-operator combination for a given lattice action to improve the accuracy e.g.~of tuning parameters for lattice simulations.

We analyze existing configurations generated by the RBC-UKQCD collaboration using 2+1 dynamical flavors of Shamir DWF and Iwasaki gauge action. We analyze these ensembles using both Wilson and Zeuthen flow and consider Wilson, clover and Symanzik operators with and without tree-level normalization. Comparing ratios of scales to estimate cutoff effects, we find that $t_0$ determined using Zeuthen flow with Symanzik operator exhibits a very consistent continuum limit and hence carries very little dependence on the lattice spacing.
Similarly, the $w_0$ scale is best identified with Zeuthen flow and clover operator. In contrast, the $t_0$ scale determined using Wilson flow and clover operator, a popular choice with Wilson gauge action, could have sizable cutoff effects on ensembles with lattice spacing $a\approx 0.08$ fm (medium ensembles). 

Frequently the ratio $\sqrt{t_0}/w_0$ is used to estimate lattice artifacts of  gradient flow scales determined with different lattice actions, flows and operators. Typically this comparison is performed at the physical point. Since we neither use external information on the lattice scale nor other quantities to perform a ``global fit'', the physical point is not accessible to us. By quoting values in the chiral limit, we obtain systematically lower results compared to determinations at the physical point as can  be inferred from Fig.~\ref{fig:chiral_extrapol}. We roughly estimate this effect to be of $O(0.005)$ i.e.~our values agree at the $1 \sigma$ level with ratios we obtain using the FLAG 2021 averages \cite{FlavourLatticeAveragingGroupFLAG:2021npn} for $\sqrt{t_0}$ and $w_0$ or the corresponding results from RBC-UKQCD \cite{RBC:2014ntl}.

Nevertheless the crude chiral extrapolation and the fact that we have insufficient data at different bare gauge coupling to better constrain the continuum limit extrapolation are obvious parts of our analysis to be improved in the future. Repeating this exercise using RBC-UKQCD's three physical point ensembles (C0, M0, F0) with M\"obius DWF kernel would e.g.~be highly interesting.

\section*{Acknowledgments}
We thank RBC-UKQCD for generating and making the 2+1 flavor gauge field ensembles available. Gradient flow measurements are calculated using Qlua \cite{Pochinsky:2008zz,qlua} on the \texttt{OMNI} cluster at the University of Siegen. The used gauge field configurations were generated on the DiRAC Blue Gene Q system at the University of Edinburgh, part of the DiRAC Facility, funded by BIS National E-infrastructure grant ST/K000411/1 and STFC grants ST/H008845/1, ST/K005804/1 and ST/K005790/1, the BG/Q computers of the RIKEN-BNL Research Center, and in part on facilities of the USQCD Collaboration, which are funded by the Office of Science of the U.S.~Department of Energy. A.H.~acknowledges support by DOE grant No.~DE-SC0010005.

{\small
  \bibliography{../General/BSM.bib}
  \bibliographystyle{JHEP-notitle}
}

\end{document}